# Incentives for Private Industrial Investment in historical perspective: the case of industrial promotion and investment promotion in Uruguay (1974-2010)

Diego Vallarino[1]

Independent Researcher

diego.vallarino@gmail.com

October 2018

**Abstract**

Using as a central instrument a new database, resulting from a compilation of historical administrative records, which covers the period 1974-2010, we can have new evidence on how industrial companies used tax benefits, and claim that these are decisive for the investment decision of the Uruguayan industrial companies during that period.

The aforementioned findings served as a raw material to also affirm that the incentives to increase investment are factors that positively influence the level of economic activity and exports, and negatively on the unemployment rate.

---

# 1. Introduction

International evidence shows that there is a (positive) correlation between investment and economic growth. Usually, without considering causality, a growth rate of the economy is related to a high rate of investment growth (Kuznets, 1973, Madison, 1983, Levine, 1992).

By way of illustration, in the case of Chile, one of the most outstanding countries in its economic performance at the end of the 20th century, Magendzo (2004) states that "the *average growth of more than 7% for the period 1990-1997 was accompanied by a growth in the capital stock of more than 6.5%, while the average growth of approximately 3.5% between 2000 and 2003 was accompanied by an average growth of 4.5% in the capital stock.*

In fact, Magendzo (2004) deepens his vision by highlighting that *"it is not strange that there is a medium and long-term correlation between GDP growth and investment , since one of the important inputs for production is physical capital and capital and The investment depends on the level of the desired product. Given the productivity of the factors and the other productive factors, the greater the desired product, the greater the capital necessary and the economic agents They will want to invest more . These two variables—capital stock and product—must move together, but it is not easy to establish which one is leading the process "* (Magendzo, 2004, p. 3).

When the causality of economic growth and investment is analyzed, specifically in gross physical capital formation, the academic advances are not as compelling as they are for the correlation. In fact, in the medium and long term there are different views on whether growth determines the highest level of investment, or if investment ends up being the determining engine of growth (for further discussion on this discussion see Barro, 1991; Blömstrom *et al* , 1993; and Barro and Sala-i-Martin, 1995). In fact, faced with this dichotomy, Magendzo (2004, p. 4) explains that *"the theoretical divergence on the role of investment in medium and long-term growth gives rise to empirical evidence playing an important role."*

Based on this background, the objective of this work is to delve deeper into the determinants of private investment in Uruguay between the years 1970 and 2010, and in particular to analyze the role played by the investment incentive mechanisms that the State established from 1974. The aim is to analyze the various dimensions that are taken into account when companies in the national industrial sector make investment decisions. Specifically, the links between the determinants of productive investment will be studied, considering the historical experience of the Uruguayan economy, in light of the most relevant contributions of economic theory and the evidence provided by national and international studies and research.

For these purposes, this work evaluates the importance of investment within the economic activity of Uruguay. It seeks to determine the effects of growth in capital investment on the level of economic activity, employment and exports, in order to justify the fiscal effort associated with incentives for investment promotion. But before developing said methodology, a small analysis will be made of the evaluation of the main variables that generally explain the investment processes.

# 2. Database Construction

The work of collecting basic information was carried out between the months of March 2009 and November 2010. During that period, 2,032 review reports on promoted projects were collected. Specifically, between 1974 and 2009, 1,384 industrial projects were declared of national interest. Specifically, information was collected on all the projects that were approved under Law 14,178, both new projects and expansions. The difference between these two concepts is that the new projects referred to companies that had no activity prior to the declaration of national interest, while the expansions corresponded to projects of companies in progress.

Continuing with the survey strategy, and once the information included in the database was completed, its processing involved the reconstruction of the tax benefits associated with each of the projects. Given that the survey was done through the review reports carried out by the UAPI or COMAP to the Executive Branch, and these being reports prior to the decree that regulated the real benefits that the government would award, in many cases the benefits were not included. explicitly. In approximately 14% of all projects and 29% of projects submitted before 1998, tax benefits were not part of the report.

In order to complete the database of projects presented and to empirically demonstrate that 100% of the projects that were recommended by the UAPI and COMAP were beneficiaries of tax exemptions, the tax benefits for each company were estimated. For this estimate, and based on the information collected, the criteria of Law 14,178 or Law 16,906 and its regulations were followed, which establishes that the benefits granted will be the equivalent of 100% of own funds or 100% of investment in fixed assets. Of these two amounts, the lower was taken, with the objective that both conditions were met at the same time.

Therefore, the database incorporated the potential tax benefit that the investor could have enjoyed, thus allowing, through economic and statistical analysis, to achieve an approximation to the extent of the tax burden of each regime. The results achieved through the estimation were contrasted with a large number of the projects for which the amount of the accessed benefit was available. They presented such coherence that it allowed us to conclude on the plausibility of the amount of potential tax benefits of all the projects presented and approved as being of national interest.

After almost two years of field surveys, working on primary information sources, the result of the empirical work involved the reconstruction of the entire database of projects declared of national interest under the Industrial Promotion Law (1974). and by the Investment Promotion Law (1998), this being the first work that allows identifying internal variables of each investment project. And this not only for the projects presented between 1974 and 2007 where the variables surveyed coincided, but through the processing of the projects approved in 2008 and 2009 it was possible to unite the database, making it comparable and continuous longitudinally. company. This will allow us to have the history of the projects presented by the companies and comparable with the projects that were presented later.

### 3. Variable Selection

This section analyzes the evolution of private investment in Uruguay and its determinants, in the period between 1970 and 2010. To do so, some of the variables included in Ribeiro and Teixeira (2007) are taken as reference, [2]where The main determining factors of private investment are analyzed for the case of Brazil during the period 1956-1996. These factors are: the product, public investment, domestic credit to the private sector, the exchange rate, the real interest rate and inflation. Additionally, the study of the evolution of tax benefits granted to the industrial sector is included, whose data arise from the database reconstructed for this work.

---

[2] "Econometric analysis of private investment in Brazil" presented in the theoretical discussion at the beginning of this work.

First of all, it is important to note that the Gross Fixed Capital Formation (FBKF) is taken as a proxy for Investment, which is composed of machinery, equipment and cultivated assets, according to data obtained from the Central Bank. From its decomposition, according to institutional sector, the data relating to Public Investment and Private Investment emerge. To carry out the econometric exercise at the end of the work (considering that the main sources were the INE and the BCU) the data are taken at constant 1983 prices.

The Real Interest Rate represents the cost of using capital for the company. For this analysis, the difference between the interest rate of the loans (nominal interest rate) and the annual average CPI (variable representative of inflation) was used as data, based on the assumption that the following relationship is met: $r = i\text{-}\pi$, where *r* represents the real interest rate, *i* represents the nominal interest rate and *π* the inflation rate.

The variation of the Inflation Rate is used as a proxy to evaluate the conditions of uncertainty in the economy, given that the latter plays an important role in stimulating investment.

The Exchange Rate variable, for its part, by representing one of the components that determine the real cost of imports, can also influence the level of investment. Taking into account that the value of the accumulated capital desired by a company is a positive function of its level of product, considering the latter as an approximation of the level of demand from the private sector, the Gross Domestic Product (GDP) is considered as another of the decisive determinants of private investment. The data referring to it were obtained from the National Accounts of the Central Bank.

Finally, given that in emerging countries a large part of companies encounter restrictions in the credit market, it is also interesting to analyze Domestic Credit to the Private Sector.

## 4. Evolution of the Investment Promotion Strategy

In the study period (1970–2010), 3 types of general investment promotion mechanisms were promoted to encourage private sector investment: Industrial Promotion Law (Law No. 14,178) in 1974; Investment Promotion Law (Law No. 16,906) in 1998 and New Investment Promotion Regime (Decree No. 455/007) in 2007.

While during the period in which the Industrial Promotion Law was in force (1974-1997), 37 projects were approved, an annual average, with the promulgation of the Investment Law (1998) - and until the regulation of the New Investment Promotion Regime (2007) - the number of approved projects became 55, annual average. With the entry into force of the New Investment Promotion Regime, there is a quantitative leap of great significance, with a number of 509 projects now being approved, also on an annual average for 2008-2010. Likewise, in 2010 a historical record was recorded with the approval of 829 projects.

Likewise, if we compare the evolution of private investment with that of the investment amounts approved under the different investment promotion mechanisms existing throughout the period 1974 - 2010, it is highlighted that, while during the period 1974 - 1997 the investment approved It represented 6% of total private investment, in the period 1998 - 2007 this relationship rose to 13%, and in the period 2008 - 2010 it rose to 23%. The above brings us closer to the initial hypothesis that the investment promotion strategies that have been implemented in our country over the last 35 years have had a clearly positive effect on total investment, increasing its participation in the total through of these mechanisms.

**5. Research Problem**

The question of the determinants of investment has already been studied for several years. However, in Uruguay, research that aims to empirically analyze what factors determine private investment is scarce . Within the studies carried out, among the factors that influence the decisions of agents in our country, the following stand out: the evolution of expected demand, access to new credits, obstacles originating in the relative cost of factors, and volatility. in the real exchange rate.

International research shows another situation, where various studies are presented in each of the countries. There is a need to establish valuations that, although limited, allow for a more in-depth study of the determinants that influence investment. In this sense, an analysis of the role of investment promotion regimes as a factor that determines private investment is not found in the background studied.

Therefore, the objective of this work is to evaluate the importance of investment in Uruguay and its determinants and, in particular, to analyze the role played by the industrial investment incentive mechanisms that the State had and has available to the private sector. since 1974. To this end, it is also intended to evaluate the importance of investment as a generator of positive externalities for the rest of the national economy, in order to justify the effort involved in implementing incentive mechanisms for the development of productive ventures.

The initial question that is intended to be answered is then the following:

*What are the determinants of private industrial investment in Uruguay?*

Taking into account the objective explained above we can reformulate it as follows:

*Can the investment promotion regimes that were and are in force in our country since 1974 be considered determinants of private industrial investment in Uruguay?*

**6 . Guiding Hypotheses**

From the study of the national and international background, the hypotheses that will guide this research are developed below, which can be thought of as answers to the questions posed above.

Based on the questions that summarize the objective of this research, two hypotheses will be formulated, the first referring to the importance of investment in Uruguay and the other regarding the role of investment promotion mechanisms as a determining factor in the development of investment. private in our country, and more specifically for the industrial sector.

The first hypothesis that will guide the empirical strategy is, regarding the factors that explained the levels of investment in our country:

*The level of economic activity, as well as the level of employment and exports of the national economy, respond positively to the increase in private investment, which justifies the incentive for the development of the latter.*

Likewise, the innovation of this study is the analysis of the influence of investment promotion, which is why it is also hypothesized that:

*Investment promotion mechanisms and the tax benefits associated with them are determinants of private industrial investment in Uruguay.*

## 7. Empirical Strategy

In this study, two lines of research are established that, although they have different objectives, are understood to be complementary to answer the questions raised. The first of these aims to study the importance of investment in the Uruguayan economy. It seeks to determine the effects of growth in capital investment on the level of economic activity, employment and exports, in order to justify the fiscal effort associated with incentives for investment promotion.

To achieve this, the work of Lorenzo (2010) is taken as a reference, in which a study is carried out that quantifies the importance of the tax incentives of different regimes in Paraguay, in terms of the opportunity costs measured by the tax waivers incurred by its implementation, and the impact of investment growth on the level of economic activity and employment.

For the second point, the research carried out by Ribeiro and Teixeira (2001) was taken as a reference, where the main factors that explain private investment in Brazil during the period 1956-1996 are analyzed. The methodology used by these authors includes the study of investment promotion regimes as determinants of private investment that the State has made available to the industrial sector since 1974.

## 8. Importance of Investment

In Lorenzo (2010) the analysis is carried out to determine the effects of investment growth on economic activity and employment, in order to justify the effort involved in implementing an incentive mechanism for the development of productive ventures. The approach involves two techniques: the first uses an Autoregressive Vector analysis to evaluate the effects of the increase in capital on GDP, and the other uses a uniquational model that studies the response of unemployment and the level of exports to variations in the investment.

### 8.1. Response of economic activity to increases in investment

As mentioned previously, with the purpose of determining the effects of investment on economic activity measured by real GDP, in Lorenzo (2010) an equation is estimated using the Vector Autoregressive (VAR) method, with the purpose of measure the dynamic effects of the forces that affect the variables considered. This technique makes it possible to observe the behavior of the variable of interest, in the face of innovations in the variables considered as sources. Innovations consist of simulated shocks of increases or decreases of one standard deviation on the sources, with the objective of determining the subsequent reactions of the variable of interest.

The VAR model considered includes four lags for each of the variables, and the cointegration equation is estimated with error correction mechanisms or VECM. The series included in the model for Paraguay, for the years 1998 – 2009, were:

- Natural logarithm of the Real Gross Domestic Product (GDP).
- Natural logarithm of the Real Gross Fixed Capital Formation (FBKF).
- Natural logarithm of the Real Effective Exchange Rate Index (ITCRE): this variable was included as a "proxy" to capture the effect of the rate of return on capital.

After performing the Johansen test and verifying the existence of a cointegration relationship between the variables of interest (FBKF, GDP and real exchange rate), a VECM was estimated with quarterly data, for the period from the second quarter of 1998 to the third quarter of 2009. The estimation of a VAR provides two statistics that allow analyzing the relationships between the variables of the system: (i) the response function of the variables to a specific shock, and (ii) the variance decomposition .

For the present study, the same procedure is applied to the case of Uruguay, to see if the same results are achieved. The VAR model also presents four lags, and the following series are included, considered annually, for the years 1970 – 2010:

- Natural logarithm of the Gross Domestic Product (GDP).
- Natural logarithm of Gross Fixed Capital Formation (FBKF).
- Natural logarithm of the Real Interest Rate (IRR), which in this case represents the cost of using capital or the cost of credit for the company.

For the objectives of this work, the most interesting thing to analyze based on the results of the described technique are the impulse-response graphs that show the evolution over time of GDP in the face of shocks of one standard deviation of investment and the Rate. of real interest.

To study the response of economic activity (GDP) to an Investment shock (FBKF), the following graph is presented where the impulse-response functions estimated based on Cholesky factorization are observed .

Based on data corresponding to the years 1970 – 2010, the response of GDP to a variation in Gross Fixed Capital Formation in 10 time periods forward. In this case, they correspond to years, since the series data are presented annually.

Until the 9th year, the response of Economic Activity to a shock in Investment presents a positive correlation, and from the last year this is null. Therefore, the response function of the GDP variable to a specific shock of the FBKF shows an effective relationship. In accordance with economic theory, in the first years the relationship between both variables is very high, decreasing after the third year. That is, if a positive shock occurs in the FBKF, the GDP will increase, and the effect of this will extend until the 9th year.

The effect of the Real Interest Rate shock on GDP is positive for the first two periods, although less significant than in the case of investment. Given the initial impulse; Economic activity grows slightly until approximately the second year, then the rise slows and after the fourth year, it begins to contract. From year four onwards, the effect fades and the relationship begins to be negative.

The results coincide with those obtained by Lorenzo for the economy of Paraguay. These confirm the impact of increased investment on GDP, and that the mechanisms that aim to increase investment make it possible to contribute to the development of economic activity.

*8.2. Response of labor and exports to increases in investment*

To evaluate the relationship between the level of employment and exports with respect to investment, an analysis is proposed, as in Lorenzo (2010), where an equation for unemployment and exports is estimated, which allows obtaining the response coefficient of these variables to changes in the capital stock. The estimated coefficients also allow simulation exercises to be carried out on the evolution of the variables, when assumptions are made that modify the capital stock.

In the model proposed for the case of Paraguay, the relationship between the unemployment rate and the variation in the capital stock between the years 1987 and 2007 is studied. In turn, lags in the unemployment rate are included as explanatory variables within the model. and two dummy variables that capture structural breaks that occurred due to the effects of the financial crises in the years 1995-1996 and 2002.

$$U = \beta_0 + \beta_1 \Delta \ln(K) + \beta_2 U_{t-1} + \beta_3 U_{t-2} + \beta_4 D96 + \beta_5 D02 + \varepsilon_t$$

Due to a possible simultaneity between the unemployment rate and the capital growth rate, the equation is estimated using the Two-Stage Least Squares (2LS) method. As the fit of the model is very satisfactory, this allows it to carry out simulations regarding the effects of increases in the capital growth rate on employment.

In this research, the relationships between unemployment - investment (Model 1) and exports - investment (Model 2) are studied for the case of Uruguay between the years 1970 and 2010, with its own model based on the one proposed for the case of Paraguay.

*Model 1: Unemployment – Investment*

Using this model, the aim is to evaluate the relationship between the unemployment rate and annual aggregate investment. The variables are the following:

*Unemployment rate* ( $U$ ). The data for the years 1970 – 1989 are obtained from the Faculty of Social Sciences, and for the remaining years (1990 – 2010) from the World Bank database ( World dataBank ).

*Unemployment rate lags* ( $U_{t-1}$ and $U_{t-2}$ ). The information comes from the same source, but these variables extend the time frame of the data back two years (1968 – 2008).

*Capital Stock Variation* ( $\Delta \ln(K)$ ). Gross capital formation is taken as a proxy, and the data series comes from the World Bank ( World dataBank ). This variable contains investment in construction, plantations and permanent crops, and machinery and equipment.

The model to be analyzed is the following:

$$U = \beta_0 + \beta_1 \Delta \ln(K) + \beta_2 U_{t-1} + \beta_3 U_{t-2} + \varepsilon_t$$

Due to a possible simultaneity between the unemployment rate and the capital growth rate, the equation was estimated using the Two-Stage Least Squares (2ELS) method. As instruments for capital growth, its lag and the GDP growth rate were used. Dummies are not included to capture structural breaks.

The results found are consistent with economic theory and with the expected results, since they show the inverse relationship between the increase in investment and the unemployment rate. The estimated coefficient of the relationship between unemployment and the rate of change in investment ( $\beta_1$ = -26.4406) is significant at 90% and negative and shows that increases in investment should contribute to reducing the unemployment rate.

*Model 2: Exports – Investment*

Using this model, the aim is to evaluate the relationship between the variation in the level of exports and the growth rate of investment added annually. The variables are the following:

*Exports* ( $X$ ). The data for the years 1970 – 1999 are obtained from the Faculty of Social Sciences, and for the remaining years (2000 – 2010) from the Uruguay XXI Institute.

*Export lags* ($X_{t-1}$ and $X_{t-2}$). The information comes from the same source, but these variables extend the time frame of the data back two years (1968 – 2008).

*Capital Stock Variation* ($\Delta \ln(K)$). Idem Model 1.

The model to be analyzed is the following:

$$X = \beta_0 + \beta_1 \Delta \ln(K) + \beta_2 X_{t-1} + \beta_3 X_{t-2} + \varepsilon_t$$

Again, given the possible simultaneity between the dependent and independent variables, the equation was estimated using the Two-Stage Least Squares (2LS) method and the instruments used are the lag of capital growth and the GDP growth rate. As in model 1, dummy variables are not included to capture structural breaks.

Model 2 shows the relationship between the increase in investment and the level of exports. As expected, the relationship between the two is positive ($\beta_1 = 4.29313$) and significant at 95%, so a higher level of investment contributes to increasing the level of exports.

Based on the results obtained in the previous section, which are presented according to the expected results and are in accordance with economic theory, simulations are carried out on the effects of increases in the investment rate on employment and exports. The fit of the models is satisfactory with the hypotheses and the theoretical framework proposed, which allows simulations to be carried out regarding the effects of increases in the capital growth rate on exports and employment.

This is done by defining two scenarios with respect to the first mentioned rate, for the period within the sample of 2000 - 2010. The first of these implies a growth of 15% of capital in 2005, and from that year on investment accelerates. at 10% per year until the end of the period. The second implies a 20% growth in investment in 2005; Then, in subsequent years it accelerates to 15% annually.

In the first scenario, it is reduced to 7.1%, which means the creation of 3,300 new jobs in the last year, since the economically active population rose to 1,665,000 people in 2010 . In the second scenario, the reduction in the unemployment rate is greater, standing at 7.01%. In this case, the reduction in the number of unemployed is 4,829 in 2010.

In the case of exports, model 2 is expressed in logarithms, however, to contrast the effects of the increase in the capital rate on exports, they are quantified in monetary terms. In 2010, the level of Uruguay's foreign sales actually observed was US$ 6,762,000,000. With the growth of investment proposed in the first scenario, they increase by 6.9% (U$S 472,318,000), and with the assumptions raised in the second they show an increase of 10.7% (U$S 720,707 ,000).

## 9. Investment determinants

Ribeiro and Teixeira (2001) evaluate the determinants of investment in Brazil based on an econometric study, and analyze the main determining factors of private investment in Brazil during the period 1956-1996. Through an empirical model used in the most recent studies on developing countries, using modern instruments that include stationarity, cointegration and exogeneity tests . The stationarity and cointegration analyzes allow us to distinguish between the short- and long-term effects of the explanatory variables.

In turn, exogeneity tests check the estimation efficiency of the model and also provide data to promote policies that encourage private sector investment. This test is carried out to avoid Lucas' criticism, which, in summary, questions the use of the estimated parameters of an econometric model to carry out policy simulations, since agents are constantly reviewing their expectations in the face of changes in the economic environment.

In this study, given that the results obtained support the use of the explanatory variables of each model as policy instruments, at least three ways of inducing an increase in private sector investments are revealed: i) increase in the level of activity economic; ii ) increase in credit and long-term financing, and iii ) increase in investments in public goods.

It is stated that during the nineties the analysis of indicators related to the product, credit and public investment variables would be sufficient to explain the fall in private investment levels in Brazil. Apart from the positive influence of the product and the negative influence of the conditions of uncertainty, in the Brazilian case it has been proven: i) the importance of long-term credits from development banks; ii ) the predominance of the crowding - in effect of private investment that public investment has, and iii ) the negative impacts of exchange rate devaluations on investment.

## *9.1. Description of the methodology*

For this research, the stages established in the previously proposed methodology are adopted, in general terms. The four steps adopted by the authors are described below:

1º) The order of integration of each of the series used is determined by applying the stationarity or unit root tests . The Augmented Dickey Fuller test (DFA) is performed .

2º) The variables and their respective lags are identified, which are significant in the industrial investment equation. The general model is estimated through autoregressive estimation. Through restriction tests , the model is gradually reduced through the elimination of variables and lags that are shown to be statistically non-significant.

Engle and Granger (1987) method is used to verify the cointegration hypothesis in the series that are shown to be integrated of order one. It is tested, on the one hand, the hypothesis of the existence of a unit root for the individual variables and, on the other hand, the hypothesis of the existence of a unit root for the residuals of the cointegrating regression . If the first hypothesis is not rejected and the second is rejected, there will be a case of cointegration of two or more time series, which suggests a long-term relationship between them.

4º) The exogeneity of the explanatory variables of the model is determined. One of the most common approaches to the study of statistical exogeneity is based on the concept of *causality in the Granger sense* . The theoretical objective of this test is to determine if a variable variable Xt and a series of lagged values of it. Once the regression has been carried out, it is analyzed whether the current and past variable To carry out the test, the use of stationary variables is required.

At this stage, the present research does not analyze the presence of superexogeneity , since, unlike the study by Ribeiro and Texeira (2001), it is not intended to establish the consequences of applying different economic policy alternatives. This research seeks to visualize the relationships that existed between the variables included in the model for the time frame analyzed (years 1975-2010), in order to find the investment determinants for said period.

*9.2. Variables included in this research*

To implement these four steps, the variables to be included are identified according to Servén and Solimano (1992) [3], who establish that there are theoretical and empirical considerations that suggest that the relevant variables to determine private investment in emerging countries are: the domestic product, the rate real interest, public investment, credit available for investment, the magnitude of the external debt, the exchange rate and macroeconomic stability. The variables included are the following:

*Private investment industrial* : the gross fixed capital formation of the industrial sector, provided by the Chamber of Industries of Uruguay, was taken as a proxy. This variable contains investment in construction, plantations and permanent crops, and machinery and equipment.

*Gross Domestic Product* : the latter being understood as an approximation of the level of demand. If this result is extended to more aggregate levels, a country's product would be considered a measure of the level of demand of the entire private sector. Annual GDP data comes from the World Bank ( World dataBank ).

*Real interest rate:* represents the cost of using capital or the cost of credit for the company. Since an increase in interest contributes to discouraging investment, a negative relationship between the two variables would be expected. For the years under study, two sources of information were combined: on the one hand , data from the National Institute of Statistics and the World Bank were used . dataBank ).

*Public investment:* There are studies that affirm that in the long term there is a crowding -in relationship between public and private investment, so this effect should be taken into account when quantifying the impact of public investment on private investment. Public capital can increase productivity by generating a positive externality, as occurs in the case of investments in infrastructure and the provision of public goods, or even with countercyclical action, raising the demand for inputs and services from the private sector. The source of information where the data on the gross formation of fixed capital developed by the public sector was obtained was the Central Bank of Uruguay (BCU).

*Real Exchange Rate* : The exchange rate can influence the level of private investment. A change in the price of this variable modifies the real acquisition costs of imported capital goods, thereby modifying the profitability of the private sector and investment may vary. Furthermore, an exchange rate devaluation in real terms can cause a reduction in the real income of the economy as a whole, also reducing the levels of activity and productive capacity desired by companies. On the other hand, the devaluation of the real exchange rate can have a positive impact on investment in sectors that produce goods tradable abroad, as it increases competitiveness and the volume of exports. The data used on the nominal exchange rate in relation to the dollar, considered annually, are obtained from the National Institute of Statistics (INE).

*Domestic Credit to the Private Sector* : given that in emerging countries a large part of companies encounter restrictions in the credit market, it is also interesting to analyze Domestic Credit to the Private Sector. The origin of the data is the World Bank ( World dataBank ).

*Inflation* : In the research carried out for Brazil, the variation in the inflation rate is used as an approximation to evaluate the conditions of uncertainty in the economy. In this study, data from the National Statistics Institute (INE) are used.

---

[3]     Extracted from Ribeiro and Teixeira (2001)

*Tax benefits associated with private industrial investment promoted through different incentive mechanisms* : The series of tax benefits granted are incorporated through the Industrial Promotion Law (Decree-Law 14,178), the Investment Promotion Law (Law 16,906) and their corresponding regulatory Decrees (Decree No. 92/998 and Decree 455/007), to industrial companies, resulting from the creation of the basis for this work.

The variable to be analyzed is private industrial investment, unlike the study by Ribeiro and Texeira, where all private investment is studied, since what is intended to be evaluated is the role of the tax benefits granted between the years 1974 and 2010 . , and between 1974 and 2007 the main objective of investment promotion was the development of the industrial sector. The external debt series is not included, due to not having data for all the years included within the time frame of the study.

All variables are expressed in thousands of pesos at constant prices from 1983, and the series extend from 1975 to 2010. The year 1975 was taken as the starting point, since, although it came into force in 1974 Law 14,178, in this first year no tax benefits associated with investment promotion were granted.

The variables mentioned above are incorporated, considered in level and in first differences.

## *9.3. Application of the methodology*

Following the methodology proposed by Ribeiro and Teixeira (2001), to analyze the main determining factors of industrial investment in Uruguay in the period between 1975 and 2010, we apply a procedure that consists of four stages:

1) Stationarity test.

2) Test of restrictions on variables and lags.

3) Cointegration test.

4) Exogeneity test.

It is important to note that throughout the econometric analysis, the natural logarithm of the series of each variable was used to stabilize their variance.

## *9.4. Stationarity test*

Dickey Fuller unit roots (DFA) test.

The test calculates the statistics for two Dickey -Fuller contrasts. In each case the null hypothesis is that the selected variable has a unit root. The first is a t-statistic based on the model:

$$(1-L)y_t = \beta_0 + (1-\alpha)y_{t-1} + \varepsilon_t$$

The null hypothesis is that **(1-α) = 0.** In the second (increased) contrast, an unrestricted regression is estimated (whose regressors are a constant, a time trend, the first lag of the variable and order lags of the first difference) and a restricted version (removing the time trend and the first lag). The contrast statistic is where T is the sample size, k is the number of parameters of the unrestricted model, and the subscripts u and r denote the unrestricted and restricted models respectively. It should be noted that the critical values for these statistics are not the usual ones; The p value is shown when it can be determined.

The results obtained in the DFA Test, for a significance level of 5%, show that the inflation rate series is integrated of zero order or stationary, while the first differences of the series of industrial investment, product, public investment, interest rate, credit and tax benefits reject the null hypothesis of non-stationarity, being, therefore, integrated of order one. In the case of the exchange rate variable, it does not achieve stationarity in first difference.

*9.5. Test of restrictions on variables and lags*

Continuing with the proposed methodology, in this second instance the variables and their respective lags are identified, which are significant in the industrial investment equation. In the first instance, the general model is estimated through autoregressive estimation. This model computes parameter estimates using the Cochrane- Orcutt generalized iterative procedure . Iterations end when two successive residual sums of squares do not differ by more than 0.005 percent or after 20 iterations. The "AR delay list" specifies the structure of the disturbance. For example, the entry "1 3 4" corresponds to the process:

$$u_t = p_1 u_{t-1} + p_3 u_{t-3} + p_4 u_{t-4} + e_t$$

This model has industrial investment as a dependent variable, which is expressed as a function of its own lags and the current and lagged values of the other variables whose series were indicated as I(1): GDP, public investment, investment rate. real interest, credit, and tax benefits.

Due to the large number of explanatory variables and the relatively small number of observations, the analysis began with the estimation of a model with two lags. This number of lags was chosen based on the Akaike information criterion, which showed a better result for this modeling.

Unlike the study by Ribeiro and Texeira, no correlation problems were detected between the incorporated series. To analyze the existence of the multicollinearity problem, the Variance Inflation Factor (VIF) mechanism was applied.

Table Model 1 shows the main results of this regression. The results of the tests for each variable show that the real interest rate, and the first differences in GDP, interest rate and credit are not significant within the period considered, so these variables are eliminated from the general model.

The next step is the estimation of the reduced model, with the inclusion only of the significant variables from Model 1. A second model is estimated omitting those variables that are not significant at 1%, that is, the restrictions to be tested are simply the exclusion of one or more regressors of the starting model. The variables that are excluded are: the constant, Interest rate (in level), and GDP in first difference. From the comparison between both models it is clear that of the 3 model selection statistics, 3 have improved.

*9.6. Cointegration tests*

For the cointegration analyses, only the integrated variables of order one that were found to be statistically significant in determining private investment were used. Through the Engle - Granger method , the aim is to find the long-term relationship between the dependent variable and those at the level that determine its behavior.

Dickey -Fuller Test on the residuals, for a model without constant and without trend, shows level stationarity. Evidence of a cointegrating relationship is found , since the hypothesis of the existence of a unit root is not rejected for the individual variables, and the hypothesis of the existence of a unit root is rejected for the residuals of the cointegrating regression .

Since the series of industrial investment, public investment, GDP, credit, and tax benefits are all I(1), the long-run equilibrium relationship is given by the following equation:

$$LnInvIndustrial = 0{,}691\,LnInvPub + 0{,}488\,LnPIB - 0{,}484\,LnCred + 0{,}198\,LnBenf + \varepsilon_t$$

Within the study period, the coefficients of the variables Public investment, GDP, and tax benefits have a positive sign, while the sign of the coefficient of the credit variable is negative. The results achieved show that, for the years analyzed, industrial investment was stimulated through investment carried out by the public sector, the level of aggregate demand, and tax benefits. The negative coefficient of the credit variable represents a restriction presented by this variable for the development of industrial investments.

### 9.7. Exogeneity test.

The concept of exogeneity is the instrument that modern Econometrics uses to face problems associated with the relative arbitrariness of the forms of specification, of the selection of exogenous variables. Compliance with the exogeneity conditions in an econometric model allows valid statistical inferences to be made and adequate economic policy projections and simulations to be obtained.

In general terms, an exogenous variable is one that is determined outside the analyzed system without implying the loss of relevant information with respect to the built model. This definition depends on the parameters of interest and the purposes of the model under consideration.

One of the most common approaches to the study of statistical exogeneity is based on the concept of *causality in the Granger sense* . From this perspective, a variable is said to be exogenous with respect to another if it is not caused (in the Granger sense ) by the latter. It is important to keep in mind that for the study of causal relationships between two economic variables, the Granger causality test requires the use of stationary variables, to avoid the risk of obtaining spurious relationships. Therefore, and given that in the previous step we have demonstrated a long-run equilibrium relationship between the series of industrial investment, public investment, GDP, credit, and tax benefits, all of which are I(1), that is, stationary in first differences, we will apply the Granger causality test on these variables.

Granger causality test allows you to quickly identify causal relationships between the explanatory variables and the variable to be explained. The theoretical objective of this test is to determine if a variable , $Y_{t-1}$, $Y_{t-2}$, $Y_{t-3}$, on the variable $X_t$ and a series of lagged values of it, that is, $X_{t-1}$, $X_{t-2}$, $X_{t-3}$, etc.

Once the regression has been carried out, it is determined whether it is easier to predict the future of the variable Y with this instrument than it would be to estimate $Y_t$ exclusively based on its past without knowing its relationship with X; In other words, it is analyzed whether the current and past variable coefficients of the regressions of Y on X as well as those of X on Y are null for the support variable, that is, that the variable provides information to explain X.

If the value of the reference statistic *"F"* exceeds the tabulated value, the null hypothesis will be rejected and therefore it will be accepted that *X* causes *Y* or vice versa. For the variables incorporated in this research, the results of the Granger Causality Test , for a significance level of 5%.

The following results emerge from the analysis:

The differential of tax benefits, credit, and public investment are not Granger causes of the industrial investment differential, which shows the presence of strict exogeneity in these three relationships. The reciprocal also holds in all three cases, so there is a bidirectional relationship between each pair of variables.

The GDP differential is a Granger cause of the industrial investment differential, which shows the absence of strict exogeneity. The reciprocal is not fulfilled; Therefore, a bidirectional relationship between both variables is not detected.

Therefore, tax benefits, credit and public investment (all expressed in first differences) are explanatory variables exogenous to the model, whose endogenous variable is the industrial investment differential. The same cannot be stated with respect to the product differential variable, which proved to be the cause of the industrial investment differential, which indicates the absence of exogeneity. It can be concluded that the inferences related to the first three parameters of the long-term model of private investment can be made without loss of relevant information, while for GDP there may be a loss of information. Therefore, a limitation is identified in the model found in the previous section.

## 10. Conclusions

This research analyzed the effects of growth in capital investment on the level of economic activity, employment and exports, in order to justify the fiscal effort associated with incentives for investment promotion. According to the methodology adopted, the response of GDP to a variation in Gross Fixed Capital Formation in 10 time periods forward was found, and it was contrasted with the effect of a shock in the Real Interest Rate. The results for the first case showed a positive effect that extends for 9 years, which is greater and more extended over time than the effect shown in the second comparison.

In turn, an equation was estimated for unemployment and exports, which allows obtaining the response coefficient of said variable to changes in the capital stock. The results found are consistent with economic theory and with the expected results, since they show the inverse relationship between the increase in investment and the unemployment rate, and a positive link between investment and exports in the Uruguayan economy for the 1970s - 2010.

Through these studies, it is concluded that the stimuli to increase private investment are factors that positively influence the level of economic activity and exports, and negatively influence the unemployment rate.

In turn, the determinants of private industrial investment in Uruguay between 1975 and 2010 were studied, with the purpose of evaluating the influence of tax incentives provided to private companies that applied the investment promotion mechanisms granted through Law 14,178, Law 16,906 and its regulatory decrees, when making the decision to invest in the industrial sector.

The results achieved showed that, for the years analyzed, industrial investment was stimulated by the level of aggregate demand, tax benefits and investment carried out by the public sector. For the first variable, a limitation is presented, since it does not detect a bidirectional relationship between GDP and industrial investment. In this sense, it can be stated that the tax exemptions granted through the aforementioned regulations favored the development of private investment in the industrial sector, and that these, in turn, contributed to reducing the unemployment rate, increasing the level of exports, and increase the level of economic activity.

**GRAPHIC ANNEX**

**Illustration 1 - FBKF of Uruguay in thousands of pesos at constant 2005 prices (1970-2010)**

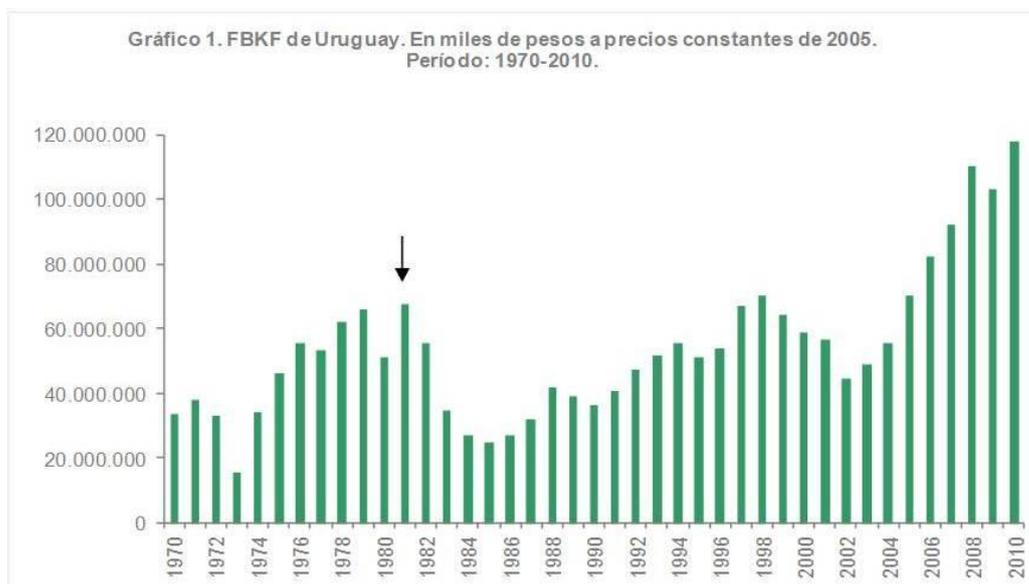

**SOURCE**: own elaboration based on data from the World Bank.

**Illustration 2 - GDP and FBKF in Uruguay 1970-2010, in thousands of constant 2005 pesos**

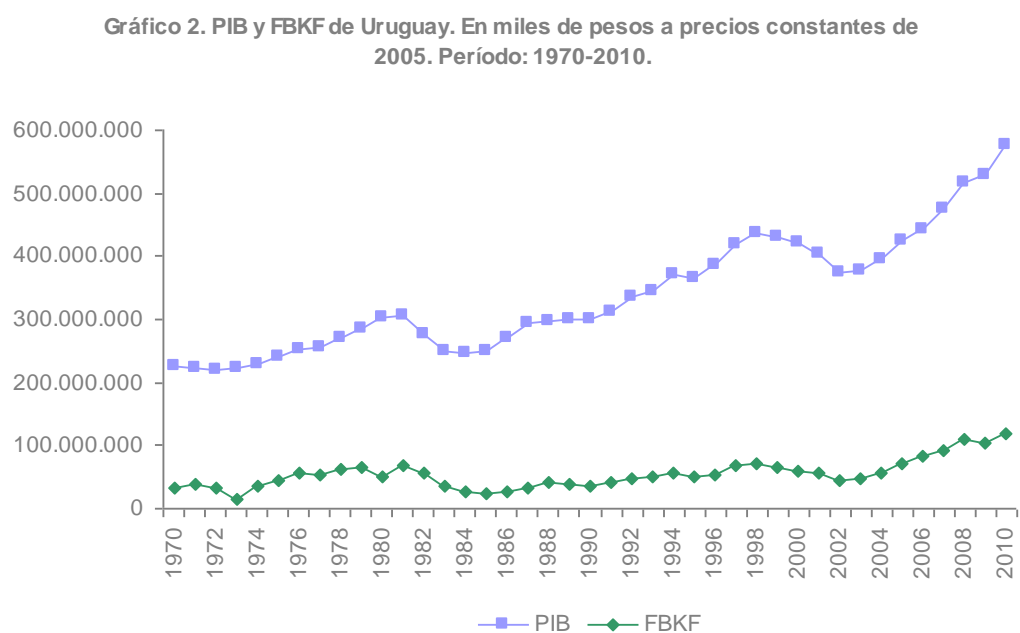

**SOURCE**: own elaboration based on data from the BCU.

**Illustration 3 - FBKF by institutional sector in Uruguay as a % of GDP (1970-2010)**

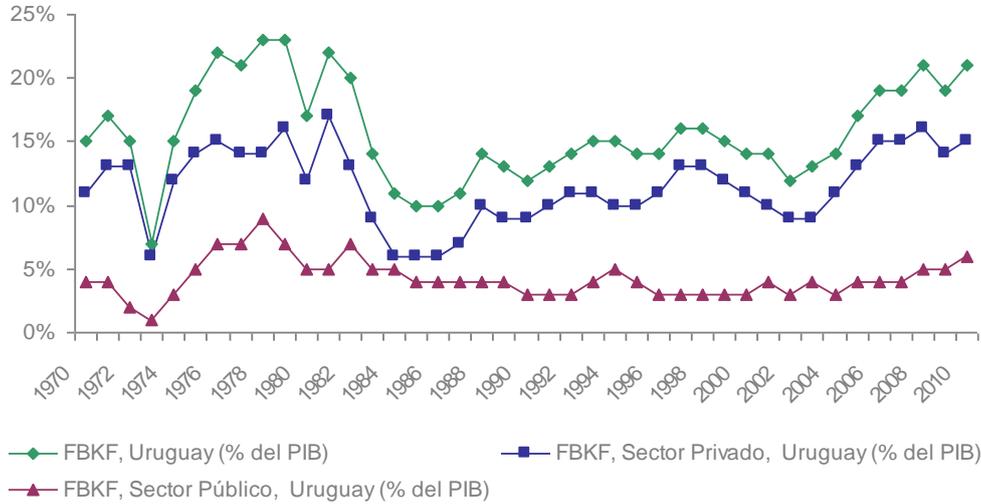

**SOURCE**: own elaboration based on data from the World Bank.

**Illustration 4 - Participation of the FBKF by institutional sector (1970-2010)**

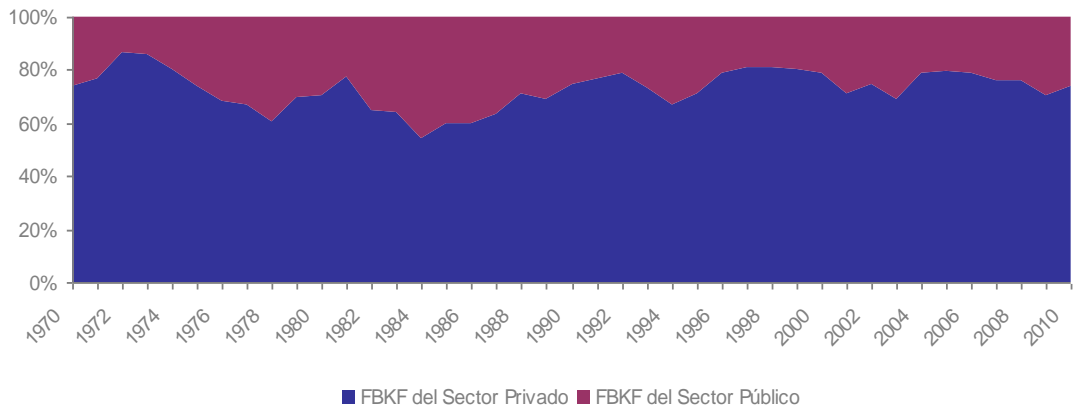

**SOURCE**: own elaboration based on data from the World Bank.

**Illustration 5 - Domestic credit in the private sector and FBKF of the private sector in thousands of constant 2005 pesos (1970-2010)**

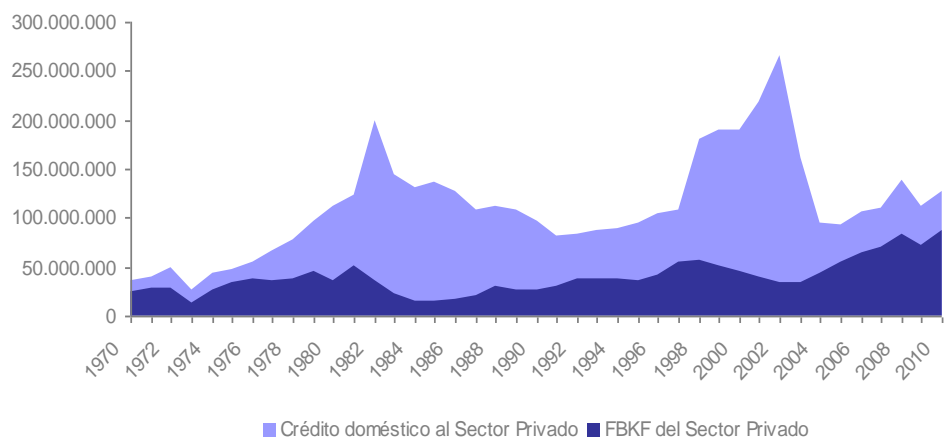

Gráfico 5. Crédito Doméstico al Sector Privado y FBKF del Sector Privado (en miles de pesos a precios constantes 2005). Período: 1970 - 2010.

**SOURCE** : own elaboration based on data from the World Bank.

**Illustration 6 - Variation of the TC and FBKF of the private sector (1970-2010)**

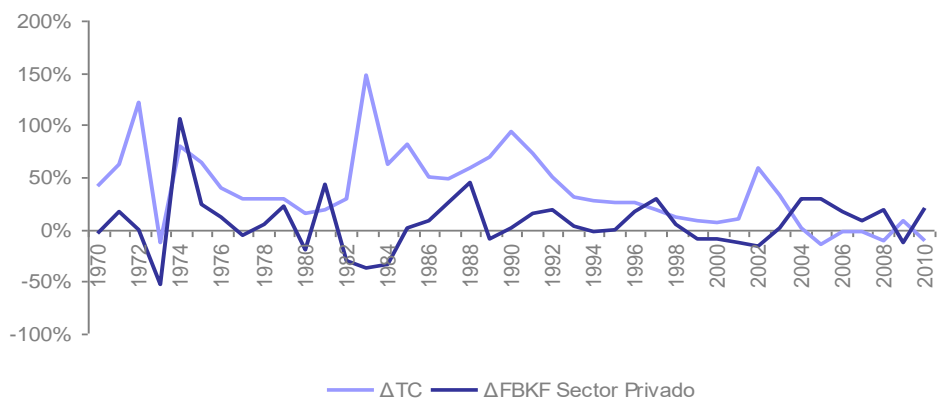

Gráfico 6. Variación del TC y de la FBKF del Sector Privado. Período 1970 - 2010.

**SOURCE** : own elaboration based on data from the World Bank.

**Illustration 7 - Variation in the interest rate and FBKF of the private sector in constant 2005 dollars (1970-2010)**

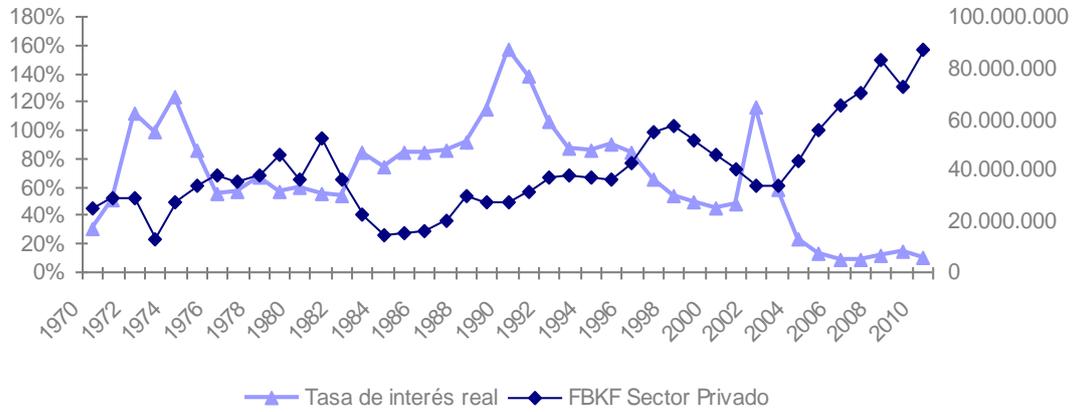

**SOURCE**: own elaboration based on data from the World Bank.

**Illustration 8 - Variation in Inflation and FBKF of the private sector in constant 2005 dollars (1970-2010)**

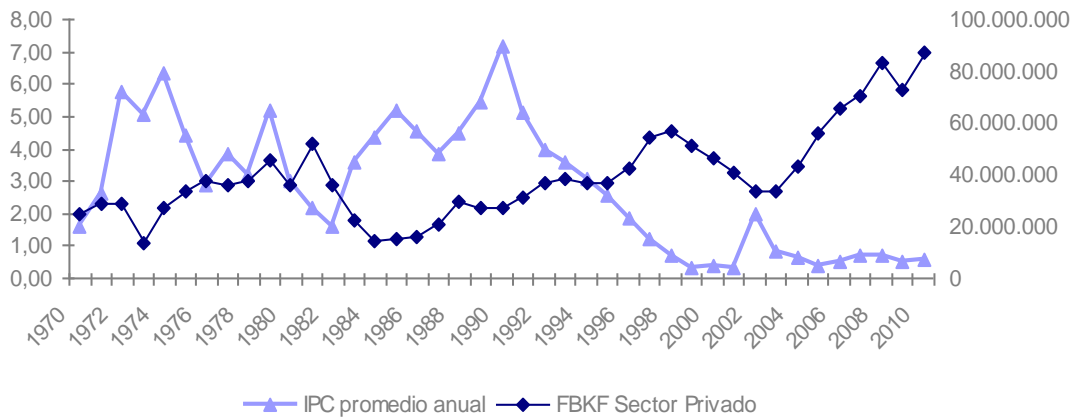

**SOURCE**: own elaboration based on data from the World Bank.

# Illustration 9 - Evolution of approved projects (1970-2010)

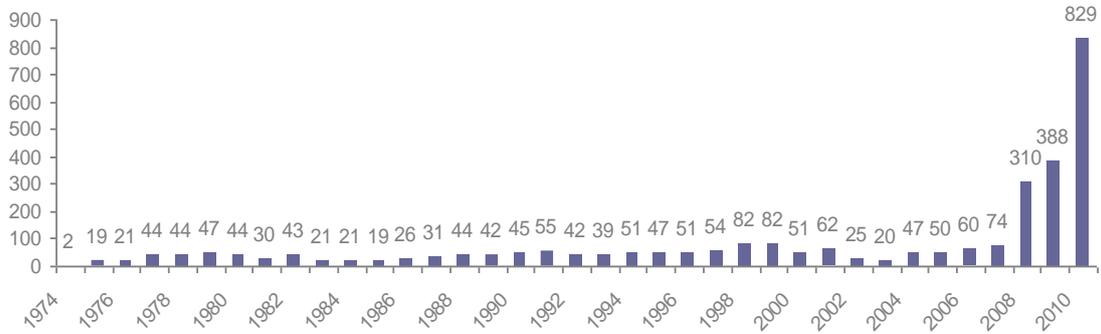

**Gráfico 9. Evolución Proyectos Aprobados.
Período 1974 - 2010.**

**SOURCE** : own elaboration based on field work.

# Illustration 10 - GDP response to a shock in the FBKF

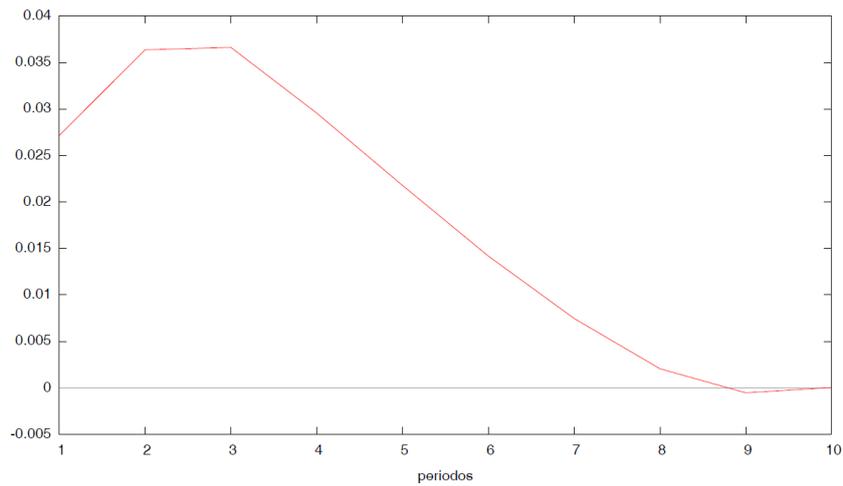

**SOURCE** : Own elaboration, based on data from the BCU, FCS, and World Bank.

**Illustration 11 - GDP response to a shock in the IRR**

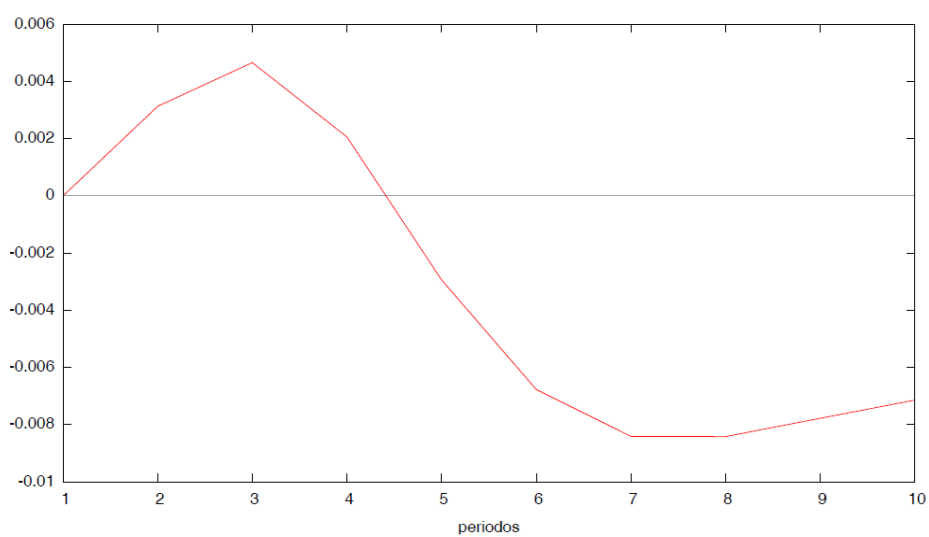

**SOURCE** : Own elaboration, based on data from the BCU, FCS, and World Bank.

*Model 1: Results of the Unemployment Equation*

| Dependent variable: *Ut* | | | | |
|---|---|---|---|---|
| **Method: Two-Stage Least Squares** | | | | |
| **Sample (adjusted): 1970 - 2010 - Observations : 41** | | | | |
| **Instruments: ΔLnGDP and ΔLnKt - 1** | | | | |
| **Variable** | **Coefficient** | **Standard error** | **Statistical - t** | **Value - p** |
| *Const* | 2.90667 | 1.01344 | 2,868 | 0.00413 |
| $\Delta \ln(K)$ | -26.4406 | 15.1637 | 15.1637 | 0.08122 |
| *Ut-1* | 1.19206 | 0.146698 | 0.146698 | 0.00001 |
| *Ut-2* | -0.479132 | 0.155168 | 0.155168 | 0.0202 |
| | | | | |
| R-squared | 0.7339 | Dependent Variable Mean | | 10.2439 |
| adjusted R-squared | 0.71242 | EE Var. Dependent | | 2.79403 |
| F-statistic | 34.0306 | Quadratic Sum of Residues | | 82.8522 |
| Prob (F-statistic) | 0.00001 | Durbin–Watson statistic | | 2.17088 |

Source: Own elaboration, based on data from the BCU, FCS, and World Bank.

*Model 2: Export Equation Results*

| Dependent variable: *x t* | | | | |
|---|---|---|---|---|
| Method: Two-Stage Least Squares | | | | |
| Sample (adjusted): 1970 – 2010. Observations : 41 | | | | |
| Instruments: ΔLnGDP and ΔLnKt - 1 | | | | |
| **Variable** | **Coefficient** | **Standard error** | **Statistical - t** | **Value - p** |
| *Const* | 0.50149 | 0.318044 | 1,577 | 0.11484 |
| $\Delta Ln\_\_(K)$ | 4.29313 | 1.34371 | 3.1950 | 0.00140 |
| *Ln x t-1* | 0.822238 | 0.164999 | 4.9833 | 0.00001 |
| *Ln x t-2* | 0.148239 | 0.161686 | 0.9168 | 0.35923 |
| | | | | |
| R-squared | 0.9809 | Dependent Variable Mean | | 14.1252 |
| adjusted R-squared | 0.979351 | EE Var. Dependent | | 0.895777 |
| F-statistic | 633,388 | Quadratic Sum of Residues | | 0.613048 |
| Prob (F-statistic) | 0.00001 | Durbin–Watson statistic | | 2.00451 |

**SOURCE** : Own elaboration, based on data from the BCU, FCS, and World Bank.

## Illustration 12 - Response of unemployment to investment

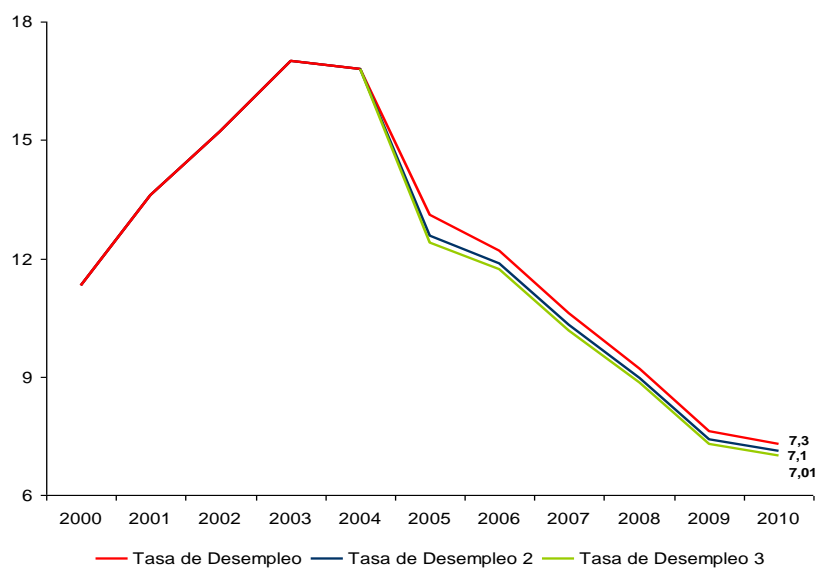

**SOURCE** : Own elaboration, based on data from the BCU, FCS, INE and World Bank.

**Illustration 13 - Response of exports to investment**

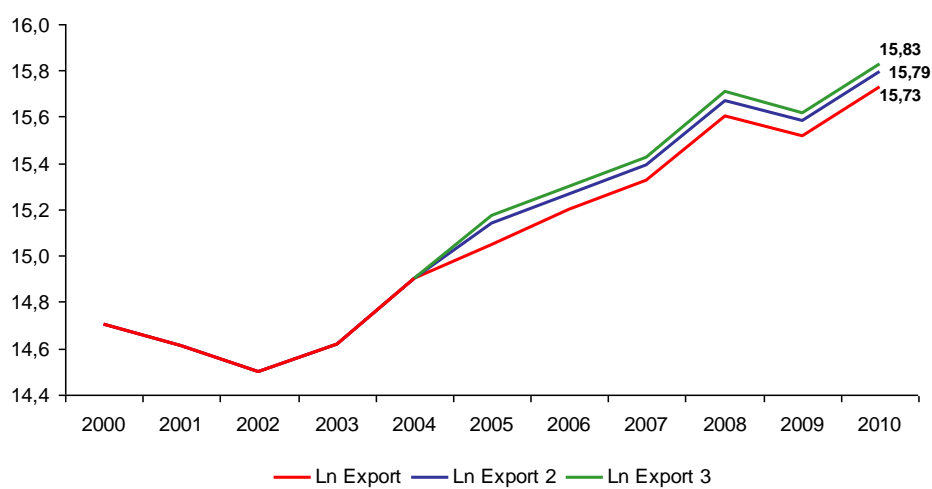

**SOURCE**: Own elaboration, based on data from the BCU, FCS, INE and World Bank.

The results for the variables analyzed are the following:

| Variable | t* DFA | Deterministic parameters |
|---|---|---|
| Log inv. industrial | -1.54248 | with constant and with tendency |
| Product log | -3.06641 | with constant and with tendency |
| Log inv. public | -1.67765 | with constant and with tendency |
| Log exchange rate | 0.16633 | with constant and with tendency |
| Log interest rate | -1.79999 | with constant and with tendency |
| Log inflation | -3.5184 | with constant and with tendency |
| Credit log | -2.77831 | with constant and with tendency |
| Log inv. benefits | -2.47475 | with constant and with tendency |
| ΔLog inv. industrial | -4,325 | with constant and without trend |
| ΔLog product | -3.28631 | with constant and without trend |
| ΔLog inv. public | -3.5002 | with constant and without trend |
| ΔLog exchange rate | -1.58168 | with constant and without trend |
| ΔLog interest rate | -4.30557 | with constant and without trend |
| ΔLog credit | -3.86923 | with constant and without trend |
| ΔLog inv. benefits | -5.40397 | with constant and without trend |

**Model 1: AR estimates using the 33 observations. Years 1978-2010**

**Dependent variable: Ln ( InvInd )**

| Variable | Coeficiente | Desv. típica | Estadístico t | valor p | |
|---|---|---|---|---|---|
| const | -3,11553 | 6,92928 | -0,4496 | 0,65807 | |
| Ln(InvPub) | 0,75413 | 0,106244 | 7,0981 | <0,00001 | *** |
| Ln(PIB) | 0,623268 | 0,20813 | 2,9946 | 0,00745 | *** |
| Ln(TasadeInt) | 0,0189578 | 0,0672379 | 0,282 | 0,78103 | |
| Ln(Inflac) | 0,0692542 | 0,065443 | 1,0582 | 0,30322 | |
| Ln(Cred) | -0,542792 | 0,178233 | -3,0454 | 0,00666 | *** |
| Ln(Benef) | 0,263858 | 0,0964936 | 2,7345 | 0,01317 | ** |
| d_Ln(InvInd) | 0,43903 | 0,0991857 | 4,4263 | 0,00029 | *** |
| d_Ln(InvPub) | -0,522098 | 0,172897 | -3,0197 | 0,00705 | *** |
| d_Ln(PIB) | 0,167003 | 0,905534 | 0,1844 | 0,85563 | |
| d_Ln(TasadeIn | 0,234664 | 0,133901 | 1,7525 | 0,09581 | * |
| d_Ln(Inflac) | -0,0744373 | 0,045267 | -1,6444 | 0,11654 | |
| d_Ln(Cred) | 0,273081 | 0,161991 | 1,6858 | 0,10819 | |
| d_Ln(Benef) | -0,0886367 | 0,059072 | -1,5005 | 0,14992 | |
| u(-1) | 0,335109 | 0,170695 | 1,9632 | 0,05865 | * |
| u(-2) | -0,2824 | 0,158365 | -1,7832 | 0,08434 | * |

| | |
|---|---|
| Suma de cuadrados de los residuos = 0,364935 | Estadístico de Durbin-Watson = 2,09817 |
| Desviación típica de los residuos = 0,13859 | Coef. de autocorr. de primer orden. = -0,051955 |
| R2 = 0,941323 | Criterio de información de Akaike = -27 |
| R2 corregido = 0,901176 | Criterio de Bayesiano de Schwarz = -6,04885 |
| Estadístico F (13, 19) = 18,8669 (valor p < 0,00001) | Criterio de Hannan-Quinn = -19,9505 |

**SOURCE** : Own elaboration

**Model 2: AR estimates using the 33 observations, years 1978-2010, with restrictions.**

**Dependent variable: Ln ( InvInd )**

| Variable | Coeficiente | Desv. típica | Estadístico t | valor p | |
|---|---|---|---|---|---|
| Ln(InvPub) | 0,782865 | 0,0728447 | 10,747 | <0,00001 | *** |
| Ln(PIB) | 0,447424 | 0,0984526 | 4,5446 | 0,00012 | *** |
| Ln(Cred) | -0,505248 | 0,0997622 | -5,0645 | 0,00003 | *** |
| Ln(Benef) | 0,312115 | 0,0395509 | 7,8915 | <0,00001 | *** |
| d_Ln(InvInd) | 0,344893 | 0,0971913 | 3,5486 | 0,00156 | *** |
| d_Ln(InvPub) | -0,487023 | 0,155833 | -3,1253 | 0,00446 | *** |
| d_Ln(TasadeIn | 0,169842 | 0,0925604 | 1,8349 | 0,07844 | * |
| d_Ln(Benef) | -0,121394 | 0,0392936 | -3,0894 | 0,00487 | *** |
| u(-1) | 0,108437 | 0,171094 | 0,6338 | 0,53087 | |
| u(-2) | -0,187896 | 0,158798 | -1,1832 | 0,24571 | |

| | |
|---|---|
| Suma de cuadrados de los residuos = 0,482434 | Estadístico de Durbin-Watson = 2,03498 |
| Desviación típica de los residuos = 0,138915 | Coef. de autocorr. de primer orden. = -0,0280207 |
| R2 = 0,922049 | Criterio de información de Akaike = -29,7889 |
| R2 corregido = 0,900222 | Criterio Bayesiano de Schwarz = -17,8168 |
| Estadístico F (8, 25) = 37,6936 (valor p < 0,00001) | Criterio de Hannan-Quinn = -25,7607 |

**SOURCE** : Own elaboration

|  | Sum of squares of the residuals | Standard deviation of the waste | Schwarz |
|---|---|---|---|
| **Complete model** | 0.364935 | 0.13859 | -6.04885 |
| **Model without var. not significant** | 0.482434 | 0.138915 | -17.8168 |

**SOURCE**: Own elaboration

### OLS estimates using the 36 observations, years 1975-2010

### Dependent variable: Ln ( InvInd )

| Variable | Coeficiente | Desv.típ. | Estad t | Valor | |
|---|---|---|---|---|---|
| Ln(InvPub) | 0,690722 | 0,106869 | 6,463 | <0,00001 | *** |
| Ln(PIB) | 0,48812 | 0,114613 | 4,259 | 0,00017 | *** |
| Ln(Cred) | -0,483606 | 0,109907 | -4,4 | 0,00011 | *** |
| Ln(Benef) | 0,198327 | 0,0512391 | 3,871 | 0,0005 | *** |

R-cuadrado = 0,999317
R-cuadrado corregido = 0,999253
Estadístico de Durbin-Watson = 2,38386
Coef. de autocorr. de primer orden. = -0,20393
Criterio de información de Akaike (AIC) = 0,918181
Criterio de Schwarz (BIC) = 7,25226
Criterio de Hannan-Quinn (HQC) = 3,12894

**SOURCE**: Own elaboration

### Dickey -Fuller test on residuals (Lag order = 1)

| Estimated value of (a - 1) | -1.22744 |
|---|---|
| Contrast statistic: tau_nc (5) | -4.40547 |
| Asymptotic P value | 0.02035 |

**SOURCE**: Own elaboration

**Results of the Granger Causality Test** (with 4 lags). Years 1975 – 2010.

| Null hypothesis (with 4 lags) | | | Obs. | F-Statistic | Prob. |
|---|---|---|---|---|---|
| dln_Invind_ _ | Granger cause. | dln_Benef_ _ | 31 | 1.62682 | 0.2030 |
| dln_Benef_ _ | Granger cause. | dln_Invind_ _ | | 0.64212 | 0.6382 |
| dln_Invind_ _ | Granger cause. | dln_Cred_ _ | 31 | 1.89825 | 0.1465 |
| dln_Cred_ _ | Granger cause. | dln_Invind_ _ | | 2.19187 | 0.1033 |
| dln_Invpub_ _ | Granger cause. | dln_Invind_ _ | 31 | 1.66061 | 0.1949 |
| dln_Invind_ _ | Granger cause. | dln_Invpub_ _ | | 2.66538 | 0.0594 |
| dln_GDP _ | Granger cause. | dln_Invind_ _ | 31 | 5.41234 | 0.0034 |
| dln_Invind_ _ | Granger cause. | dln_GDP _ | | 3.61315 | 0.0207 |

**SOURCE**: Own elaboration

**Model: OLS estimates using the 41 observations 1970-2010.**

**Dependent variable: Ln ( InvInd )**

| *Variable* | *Coeficiente* | *Desv. típica* | *Estadístico t* | *valor p* | |
|---|---|---|---|---|---|
| **Ln(InvPub)** | 0,680912 | 0,11968 | 5,6894 | <0,00001 | *** |
| **Ln(PIB)** | 0,560538 | 0,107406 | 5,2189 | <0,00001 | *** |
| **Ln(Cred)** | -0,501532 | 0,115802 | -4,3309 | 0,0001 | *** |

| | |
|---|---|
| Desviación típica de la var. dependiente. = 0,431553 | Estadístico de Durbin-Watson = 1,43048 |
| Suma de cuadrados de los residuos = 3,21856 | Coef. de autocorr. de primer orden. = 0,274061 |
| Desviación típica de los residuos = 0,291031 | Log-verosimilitud = -6,0114 |
| R2=0,998882 | Criterio de información de Akaike = 18,0228 |
| R2 corregido = 0,998823 | Criterio de Schwarz = 23,1635 |
| Estadístico F (3, 38) = 11314,6 (valor p < 0,00001) | Criterio de Hannan-Quinn = 19,8948 |

**SOURCE**: Own elaboration

Gretl econometric program, for the established model we obtain the following results:

For the change point located in the year 1974

| **Chow test of structural change in observation 1974** |
|---|
| Null hypothesis: there is no structural change |
| Contrast statistic: $F(3, 35) = 1.78361$ |
| with p value = $P(F(3, 35) > 1.78361) = 0.168266$ |

For the change point located in the year 1998

| **Chow test of structural change in observation 1998** |
|---|
| Null hypothesis: there is no structural change |
| Contrast statistic: F(3, 35) = 3.16873 |
| with p value = P(F(3, 35) > 3.16873) = 0.036302 |

**SOURCE** : Own elaboration

For the change point located in the year 2007

| **Chow contrast of structural change in the 2007 observation** |
|---|
| **Null hypothesis: there is no structural change** |
| **Contrast statistic: F(3, 35) = 3.76652** |
| **with p value = P(F(3, 35) > 3.76652) = 0.0191929** |

**SOURCE** : Own elaboration

**Illustration 14 - Regression residual (Observed – estimated industrial investment)**

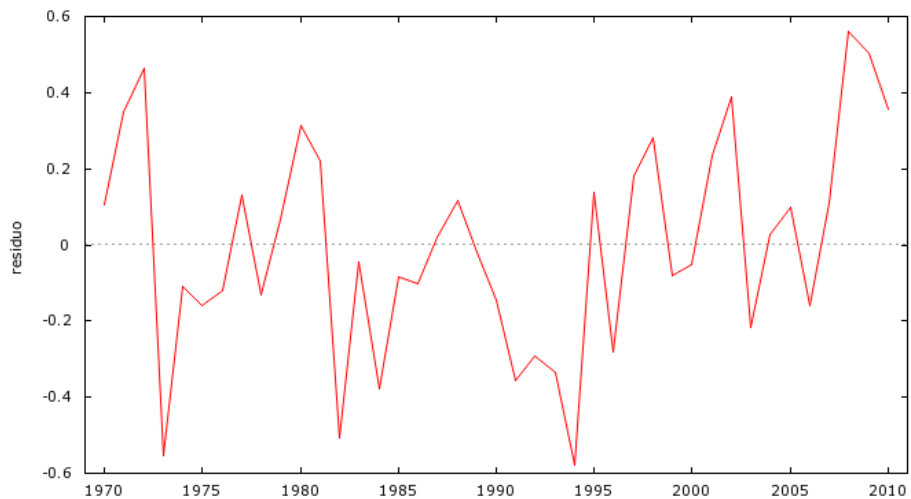

**SOURCE** : Own elaboration

**INFORMATION ANNEX.**

**Gross Fixed Capital Formation.**

**Years 1955 – 2010, in thousands of pesos at constant prices of the year 1983.**

| Year | Private investment | Public Investment | Total inversion |
|---|---|---|---|
| 1955 | 20,799 | 5,546 | 26,345 |
| 1956 | 18,942 | 5,387 | 24,329 |
| 1957 | 20,329 | 4,630 | 24,959 |
| 1958 | 14,236 | 3,849 | 18,085 |
| 1959 | 15,513 | 3,488 | 19,001 |
| 1960 | 17,227 | 3,681 | 20,908 |
| 1961 | 18,634 | 4,501 | 23,135 |
| 1962 | 19,105 | 4,509 | 23,614 |
| 1963 | 14,936 | 4,619 | 19,555 |
| 1964 | 14,052 | 3,167 | 17,219 |
| 1965 | 13,663 | 2,733 | 16,396 |
| 1966 | 13,195 | 2,932 | 16,127 |
| 1967 | 14,886 | 3,308 | 18,194 |
| 1968 | 13,335 | 3,556 | 16,891 |
| 1969 | 17,038 | 4,543 | 21,581 |
| 1970 | 16,962 | 6,106 | 23,068 |
| 1971 | 16,477 | 6,316 | 22,793 |
| 1972 | 14,851 | 4,471 | 19,322 |
| 1973 | 12,727 | 3,625 | 16,352 |
| 1974 | 13,569 | 4,539 | 18,108 |
| 1975 | 16,482 | 8,933 | 25,415 |
| 1976 | 18,861 | 13,755 | 32,616 |
| 1977 | 18,796 | 16,166 | 34,962 |
| 1978 | 20,066 | 19,929 | 39,995 |
| 1979 | 28,756 | 18,837 | 47,593 |
| 1980 | 34,890 | 15,721 | 50,611 |
| 1981 | 34,687 | 14,402 | 49,089 |
| 1982 | 22,849 | 18,942 | 41,791 |
| 1983 | 17,603 | 10,482 | 28,085 |
| 1984 | 12,485 | 9,162 | 21,647 |
| 1985 | 10,628 | 6,715 | 17,343 |
| 1986 | 11,723 | 7,845 | 19,568 |
| 1987 | 15,447 | 9,468 | 24,915 |
| 1988 | 17,877 | 8,140 | 26,017 |
| 1989 | 16,466 | 8,329 | 24,795 |
| 1990 | 17,053 | 5,889 | 22,942 |
| 1991 | 18,964 | 8,855 | 27,819 |
| 1992 | 24,861 | 7,884 | 32,745 |
| 1993 | 26,572 | 10,904 | 37,476 |
| 1994 | 27,560 | 12,291 | 39,851 |
| nineteen ninety five | 29,075 | 8,794 | 37,869 |
| nineteen ninety six | 32,753 | 8,970 | 41,723 |
| 1997 | 36,778 | 9,181 | 45,959 |
| 1998 | 39,868 | 9,631 | 49,499 |
| 1999 | 34,303 | 11,190 | 45,493 |
| 2000 | 29,342 | 10,200 | 39,542 |
| 2001 | 26,759 | 9,061 | 35,820 |
| 2002 | 18,013 | 6,168 | 24,181 |
| 2003 | 15,950 | 5,482 | 21,432 |
| 2004 | 22,387 | 5,528 | 27,915 |
| 2005 | 28,077 | 6,478 | 34,555 |
| 2006 | 34,074 | 7,978 | 42,052 |
| 2007 | 36,103 | 8,563 | 44,666 |
| 2008 | 45,498 | 11,285 | 56,783 |
| 2009 | 39,716 | 14,800 | 54,516 |
| 2010 | 47,608 | 14,998 | 62,371 |

Source: own elaboration based on data from the Central Bank of Uruguay.

**Gross Domestic Product (GDP), in pesos at constant 1983 prices.**

**Years 1960 - 2010.**

| Years | GDP in pesos at constant 1983 prices |
|---|---|
| 1960 | 195,871,088,205 |
| 1961 | 200,754,228,745 |
| 1962 | 197,593,255,298 |
| 1963 | 197,930,070,493 |
| 1964 | 202,760,594,989 |
| 1965 | 204,880,667,958 |
| 1966 | 211,155,867,692 |
| 1967 | 203,434,247,914 |
| 1968 | 207,276,941,296 |
| 1969 | 219,433,196,286 |
| 1970 | 224,553,041,044 |
| 1971 | 223,987,815,113 |
| 1972 | 221,032,139,412 |
| 1973 | 221,640,418,953 |
| 1974 | 228,057,672,943 |
| 1975 | 241,962,153,789 |
| 1976 | 251,484,502,169 |
| 1977 | 255,144,761,541 |
| 1978 | 268,856,203,592 |
| 1979 | 285,522,913,468 |
| 1980 | 302.207.162.026 |
| 1981 | 306,920,263,460 |
| 1982 | 276,971,378,692 |
| 1983 | 248,514,233,559 |
| 1984 | 245,674,666,648 |
| 1985 | 249,277,574,079 |

| Year | Value |
|------|-------|
| 1986 | 271,238,450,517 |
| 1987 | 292,918,911,990 |
| 1988 | 297,256,857,855 |
| 1989 | 300,538,279,442 |
| 1990 | 301,431,925,114 |
| 1991 | 312,099,023,685 |
| 1992 | 336,853,433,660 |
| 1993 | 345,805,468,985 |
| 1994 | 370,984,750,119 |
| 1995 | 365,614,378,667 |
| 1996 | 386,008,194,310 |
| 1997 | 419,002,951,966 |
| 1998 | 437,937,234,847 |
| 1999 | 429,444,702,756 |
| 2000 | 421,156,717,892 |
| 2001 | 404,966,906,333 |
| 2002 | 373,654,835,899 |
| 2003 | 376,663,818,208 |
| 2004 | 395,512,679,678 |
| 2005 | 425,018,447,915 |
| 2006 | 443,401,883,396 |
| 2007 | 475,922,428,263 |
| 2008 | 516,834,412,576 |
| 2009 | 530,175,852,170 |
| 2010 | 575,069,421,701 |

Source: Central Bank of Uruguay (BCU)

**Real interest rate. Years 1970 - 2010**

| Year | Real interest rate |
|---|---|
| 1970 | 31,208 |
| 1971 | 50,562 |
| 1972 | 112,239 |
| 1973 | 98,789 |
| 1974 | 123,476 |
| 1975 | 85,178 |
| 1976 | 55,356 |
| 1977 | 56,365 |
| 1978 | 66,462 |
| 1979 | 56,371 |
| 1980 | 59,530 |
| 1981 | 54,544 |
| 1982 | 53,377 |
| 1983 | 84,337 |
| 1984 | 73,822 |
| 1985 | 83,629 |
| 1986 | 84,394 |
| 1987 | 86,097 |
| 1988 | 90,839 |
| 1989 | 114,335 |
| 1990 | 156,629 |
| 1991 | 138,453 |
| 1992 | 106,640 |
| 1993 | 87,794 |
| 1994 | 86,189 |
| 1995 | 90,492 |
| 1996 | 84,101 |
| 1997 | 66,001 |
| 1998 | 53,698 |
| 1999 | 49,686 |
| 2000 | 45,645 |
| 2001 | 48,261 |
| 2002 | 116,425 |
| 2003 | 58,129 |
| 2004 | 23,065 |
| 2005 | 13,208 |
| 2006 | 8,732 |
| 2007 | 8,256 |
| 2008 | 11,712 |
| 2009 | 14,804 |
| 2010 | 9,438 |

Source: Own elaboration based on data from the World Bank and National Institute of Statistics (INE).

**Consumer Price Index. Years 1970 - 2010**

| Years | Annual average CPI | Δ CPI |
|---|---|---|
| 1970 | 1.6018 | 0.4053 |
| 1971 | 2.5984 | 0.6222 |
| 1972 | 5.7713 | 1.2211 |
| 1973 | 5.0609 | 0.1231 |
| 1974 | 6.3337 | 0.2515 |
| 1975 | 4.3922 | 0.3065 |
| 1976 | 2.8619 | 0.3484 |
| 1977 | 3.8556 | 0.3472 |
| 1978 | 3.2111 | 0.1672 |
| 1979 | 5.1879 | 0.6156 |
| 1980 | 3.0231 | 0.4173 |
| 1981 | 2.1718 | 0.2816 |
| 1982 | 1.5931 | 0.2665 |
| 1983 | 3.5922 | 1.2548 |
| 1984 | 4.3335 | 0.2064 |
| 1985 | 5.1843 | 0.1963 |
| 1986 | 4.5596 | 0.1205 |
| 1987 | 3.8587 | 0.1537 |
| 1988 | 4.4850 | 0.1623 |
| 1989 | 5.4653 | 0.2186 |
| 1990 | 7.1782 | 0.3134 |
| 1991 | 5.1032 | 0.2891 |
| 1992 | 3.9426 | 0.2274 |
| 1993 | 3.6018 | 0.0864 |
| 1994 | 3.0937 | 0.1411 |
| 1995 | 2.5625 | 0.1717 |
| 1996 | 1.8330 | 0.2847 |
| 1997 | 1.1842 | 0.3540 |
| 1998 | 0.6930 | 0.4148 |
| 1999 | 0.3415 | 0.5072 |
| 2000 | 0.4126 | 0.2084 |
| 2001 | 0.2950 | 0.2851 |
| 2002 | 1.9550 | 5.6271 |
| 2003 | 0.8117 | 0.5848 |
| 2004 | 0.6142 | 0.2433 |
| 2005 | 0.4017 | 0.3460 |
| 2006 | 0.5183 | 0.2905 |
| 2007 | 0.6833 | 0.3183 |
| 2008 | 0.7358 | 0.0768 |
| 2009 | 0.4808 | 0.3465 |
| 2010 | 0.5617 | 0.1681 |

Source: National Institute of Statistics (INE).

**Dollar exchange rate. Years 1970 - 2010**

| Year | Annual average TC | ΔTC |
|------|-------------------|---------|
| 1970 | 0.00028 | 0.40529 |
| 1971 | 0.00046 | 0.62000 |
| 1972 | 0.00102 | 1.21739 |
| 1973 | 0.00089 | -0.12432 |
| 1974 | 0.00161 | 0.79884 |
| 1975 | 0.00263 | 0.63766 |
| 1976 | 0.00368 | 0.39952 |
| 1977 | 0.00472 | 0.28253 |
| 1978 | 0.00608 | 0.28805 |
| 1979 | 0.00785 | 0.29048 |
| 1980 | 0.00910 | 0.15854 |
| 1981 | 0.01081 | 0.18811 |
| 1982 | 0.01391 | 0.28761 |
| 1983 | 0.03447 | 1.47735 |
| 1984 | 0.05600 | 0.62469 |
| 1985 | 0.10133 | 0.80941 |
| 1986 | 0.15174 | 0.49748 |
| 1987 | 0.22620 | 0.49069 |
| 1988 | 0.35894 | 0.58684 |
| 1989 | 0.60501 | 0.68556 |
| 1990 | 1.17027 | 0.93429 |
| 1991 | 2.01824 | 0.72460 |
| 1992 | 3.02586 | 0.49926 |
| 1993 | 3.94601 | 0.30410 |
| 1994 | 5.05013 | 0.27981 |
| 1995 | 6.34721 | 0.25684 |
| 1996 | 7.96996 | 0.25566 |
| 1997 | 9.44250 | 0.18476 |
| 1998 | 10.47063 | 0.10888 |
| 1999 | 11.33671 | 0.08272 |
| 2000 | 12.09715 | 0.06708 |
| 2001 | 13.31645 | 0.10079 |
| 2002 | 21.23545 | 0.59468 |
| 2003 | 28.18361 | 0.32720 |
| 2004 | 28.68066 | 0.01764 |
| 2005 | 24.45358 | -0.14738 |
| 2006 | 24.04837 | -0.01657 |
| 2007 | 23.44604 | -0.02505 |
| 2008 | 20.94932 | -0.10649 |
| 2009 | 22.56797 | 0.07726 |
| 2010 | 20.05924 | -0.11116 |

Source: National Institute of Statistics (INE).

**Unemployment rate. Years 1968 - 2010**

| Year | Unemployment rate |
|------|-------------------|
| 1968 | 8.6 |
| 1969 | 8.9 |
| 1970 | 7.4 |
| 1971 | 7.8 |
| 1972 | 7.8 |
| 1973 | 8.9 |
| 1974 | 7.9 |
| 1975 | 11.2 |
| 1976 | 12.3 |
| 1977 | 11.4 |
| 1978 | 10.1 |
| 1979 | 8.1 |
| 1980 | 6.4 |
| 1981 | 7.3 |
| 1982 | 6.6 |
| 1983 | 11.7 |
| 1984 | 15.4 |
| 1985 | 13.9 |
| 1986 | 13 |
| 1987 | 10.7 |
| 1988 | 9.1 |
| 1989 | 8.6 |
| 1990 | 8 |
| 1991 | 8.5 |
| 1992 | 9 |
| 1993 | 9 |
| 1994 | 8.3 |

| Year | Value |
|------|------|
| 1995 | 9.2 |
| 1996 | 10.2 |
| 1997 | 11.9 |
| 1998 | 11.4 |
| 1999 | 10 |
| 2000 | 11.3 |
| 2001 | 13.6 |
| 2002 | 15.2 |
| 2003 | 17 |
| 2004 | 16.8 |
| 2005 | 13.1 |
| 2006 | 12.2 |
| 2007 | 10.6 |
| 2008 | 9.2 |
| 2009 | 7.6 |
| 2010 | 7.3 |

Source: National Institute of Statistics (INE).

**Exports, in thousands of dollars. Years 1970 - 2010**

| Years | Exports |
|---|---|
| 1970 | 232,709 |
| 1971 | 205,693 |
| 1972 | 214,077 |
| 1973 | 321,510 |
| 1974 | 382,182 |
| 1975 | 383,847 |
| 1976 | 546,476 |
| 1977 | 607,523 |
| 1978 | 686,053 |
| 1979 | 788,134 |
| 1980 | 1,058,549 |
| 1981 | 1,215,377 |
| 1982 | 1,022,884 |
| 1983 | 1,045,148 |
| 1984 | 924,584 |
| 1985 | 853,611 |
| 1986 | 1,087,823 |
| 1987 | 1,182,323 |
| 1988 | 1,404,510 |
| 1989 | 1,598,775 |
| 1990 | 1,692,927 |
| 1991 | 1,604,724 |
| 1992 | 1,702,501 |
| 1993 | 1,645,312 |
| 1994 | 1,913,432 |
| 1995 | 2,105,957 |
| 1996 | 2,397,224 |

| Year | Value |
|------|-------|
| 1997 | 2,725,741 |
| 1998 | 2,770,654 |
| 1999 | 2,237,114 |
| 2000 | 2,427,038 |
| 2001 | 2,209,674 |
| 2002 | 1,977,445 |
| 2003 | 2,232,346 |
| 2004 | 2,970,437 |
| 2005 | 3,423,724 |
| 2006 | 3,989,402 |
| 2007 | 4,516,409 |
| 2008 | 5,988,892 |
| 2009 | 5,497,000 |
| 2010 | 6,762,000 |

Source: own elaboration based on data from Uruguay XXI and Faculty of Social Sciences

**Industrial Investment, in thousands of pesos at constant 1983 prices .**

**Years 1970 - 2010**

| Year | Industrial Investment |
|------|----------------------|
| 1970 | 4,936 |
| 1971 | 6,063 |
| 1972 | 4,862 |
| 1973 | 2,062 |
| 1974 | 2,981 |
| 1975 | 4,398 |
| 1976 | 5,870 |
| 1977 | 7,742 |
| 1978 | 6,539 |
| 1979 | 7,084 |
| 1980 | 7,727 |
| 1981 | 6,367 |
| 1982 | 2,752 |
| 1983 | 3,232 |
| 1984 | 2,207 |
| 1985 | 2,357 |
| 1986 | 2,797 |
| 1987 | 4,067 |
| 1988 | 3,995 |
| 1989 | 3,638 |
| 1990 | 2,689 |
| 1991 | 3,195 |
| 1992 | 3,227 |
| 1993 | 3,860 |
| 1994 | 3,363 |
| 1995 | 5,268 |

| Year | Value |
|------|-------|
| 1996 | 3,450 |
| 1997 | 5,700 |
| 1998 | 5,195 |
| 1999 | 3,859 |
| 2000 | 3,692 |
| 2001 | 4,124 |
| 2002 | 3,214 |
| 2003 | 2,086 |
| 2004 | 3,586 |
| 2005 | 4,515 |
| 2006 | 3,858 |
| 2007 | 5,425 |
| 2008 | 9,520 |
| 2009 | 12,295 |
| 2010 | 10,517 |

Source: own elaboration based on data from the Chamber of Industries of Uruguay (CIU)

**Domestic credit to the Private Sector, as a percentage of GDP.**

**Years 1970 - 2010**

| Year | Domestic credit to the Private Sector, as a percentage of GDP. |
|------|---------------------------------|
| 1970 | 10% |
| 1971 | 12% |
| 1972 | twenty-one% |
| 1973 | -Four. Five% |
| 1974 | 63% |
| 1975 | 12% |
| 1976 | 14% |
| 1977 | twenty% |
| 1978 | 18% |
| 1979 | 25% |
| 1980 | fifteen% |
| 1981 | 10% |
| 1982 | 62% |
| 1983 | -28% |
| 1984 | -10% |
| 1985 | 5% |
| 1986 | -7% |
| 1987 | -fifteen% |
| 1988 | 4% |
| 1989 | -4% |
| 1990 | -eleven% |
| 1991 | -16% |
| 1992 | 4% |
| 1993 | 3% |
| 1994 | 3% |

| Year | Value |
|------|-------|
| 1995 | 7% |
| 1996 | 10% |
| 1997 | 5% |
| 1998 | 65% |
| 1999 | 5% |
| 2000 | 0% |
| 2001 | fifteen% |
| 2002 | twenty-one% |
| 2003 | -39% |
| 2004 | -41% |
| 2005 | -1% |
| 2006 | 14% |
| 2007 | 3% |
| 2008 | 27% |
| 2009 | -twenty% |
| 2010 | 14% |

Source: World Bank